# Ion Intercalation in Lanthanum Strontium Ferrite for Aqueous Electrochemical Energy Storage Devices

Yunqing Tang[1], Francesco Chiabrera[1,2]*, Alex Morata[1], Andrea Cavallaro[3], Maciej O. Liedke[4], Hemesh Avireddy [1], Mar Maller[1], Maik Butterling[4], Andreas Wagner[4], Michel Stchakovsky[5], Federico Baiutti[1,6], Ainara Aguadero[3,7], Albert Tarancón[1,8]*.

[1]Department of Advanced Materials for Energy Applications, Catalonia Institute for Energy Research (IREC), Jardins de les Dones de Negre 1, 08930, Sant Adrià del Besòs, Barcelona, Spain.

[2]Department of Energy Conversion and Storage, Functional Oxides group, Technical University of Denmark, Fysikvej, 310, 233, 2800 Kgs. Lyngby, Denmark.

[3]Department of Materials, Imperial College London, SW7 2AZ, London, UK.

[4]Helmholtz-Zentrum Dresden-Rossendorf, Institute of Radiation Physics, 01328 Dresden, Germany.

[5]HORIBA Scientific, 14, Boulevard Thomas Gobert, Passage Jobin Yvon, CS 45002-91120 Palaiseau-France.

[6]Department of Materials Chemistry, National Institute of Chemistry, Hajdrihova 19, Ljubljana SI-1000, Slovenia.

[7]Instituto de Ciencia de Materiales de Madrid, ICMM-CSIC, Sor Juana Ines de la Cruz, 3, 28049, Madrid, Spain.

[8]ICREA, Passeig Lluís Companys 23, 08010, Barcelona, Spain.

ABSTRACT

Ion intercalation of perovskite oxides in liquid electrolytes is a very promising method for controlling their functional properties while storing charge, which opens the potential application in different energy and information technologies. Although the role of defect chemistry in the oxygen intercalation in a gaseous environment is well established, the mechanism of ion intercalation in liquid electrolytes at room temperature is poorly understood. In this study, the defect chemistry during ion intercalation of $La_{0.5}Sr_{0.5}FeO_{3-\delta}$ thin films in alkaline electrolytes is studied. Oxygen and proton intercalation into the LSF perovskite structure is observed at moderate electrochemical potentials (0.5 V to -0.4 V), giving rise to a change in the oxidation state of Fe (as a charge compensation mechanism). The variation of the concentration of holes as a function of the intercalation potential was characterized by in-situ ellipsometry and the concentration of electron holes was indirectly quantified for different electrochemical potentials. Finally, a dilute defect chemistry model that describes the variation of defect species during ionic intercalation was developed.

## INTRODUCTION

Transition metal oxides (TMOs) are promising electrode materials for room temperature energy conversion and storage applications based on ion intercalation, such as ion batteries,[1–3] supercapacitors [4,5] and pseudocapacitors.[6,7] In particular, perovskite oxides such as $La_{1-x}Sr_xMnO_{3-\delta}$ (LSM), $La_{1-x}Sr_xCoO_{3-\delta}$ (LSC) and $La_{1-x}Sr_xFeO_{3-\delta}$ (LSF) are able to reversibly intercalate oxygen at room temperature in different liquid electrolytes, which raised the attention for their application as high energy density pseudocapacitors,[8–10] efficient catalyst in oxygen reduction reaction (ORR) and oxygen evolution reaction (OER),[11–13] rapid and efficient electrochromic windows and synaptic memories.[14–16] In these materials, ion intercalation is a complex phenomenon that involves redox reactions, ionic and electronic transfer through a liquid-solid interface and ionic transport.[16] Ion intercalation in perovskites oxides usually involves a change in the oxidation state of the B-site metal, which results in strong modifications of the electronic band structure of the oxide affecting its properties (electronic, thermal and optical conductivity etc.)[5,10,15,17–19] The oxygen anion exchange between the electrolyte and the perovskite may also induces a topotactic phase transformation between the oxidized perovskite and reduced brownmillerite phase,[20,21] favoring applications in resistive-switching memories and magnetoelectric and spintronic devices.[22,23]

A representative example of a pseudocapacitive behavior depending on point defects is found for non-stoichiometric $La_{0.5}Sr_{0.5}FeO_{3-\delta}$ (LSF50).[10] LSF50 crystallizes in a perovskite structure where the Sr-substitution of La creates oxygen vacancies and/or electron holes to maintain electroneutrality. Moreover, insertion (or extraction) of oxygen anions into (or from) LSF50 generates changes in the concentration of $Fe^{4+}$ holes and oxygen vacancies.[10] The increased concentration of oxygen vacancies in LSF not only causes a fast oxygen diffusion rate and an increased oxygen-anion-based pseudocapacitance,[10] but it also modifies the electrochemical performance of the material.[24,25] Apart from the straightforward application in supercapacitors, the ability to intercalate ions at room temperature for LSF family was found to offer interesting application as rocking oxygen battery and synaptic memories.[26,27]

Despite this significant interest in low temperature ion intercalation in perovskite oxides, the exact mechanism and dynamics of ionic insertion are still unclear. For instance, different works show that the main species diffusing in and out LSF electrodes during intercalation in alkaline electrolytes are oxygen species,[5,7,10,28] while others proposed that protons are the main responsible for its pseudocapacitive behavior.[29] Moreover, ion intercalation in these materials is usually interpreted on the basis of pseudocapacitive Nernst behavior, where the effect of the different point defects is not taken into account.[30] In this sense, a comprehensive defect chemistry model considering both the intercalating species and the point defects in the perovskite oxide is still missing.

The common intercalation mechanism described in literature for explaining the electrochemical reactions in perovskite oxides cycled in alkaline electrolytes involves the incorporation of oxygen species through the annihilation of an oxygen vacancy ($V_O^{\bullet\bullet}$) and the formation of two localized electronic holes ($Fe_{Fe}^{\bullet}$), as:[10,26,30]

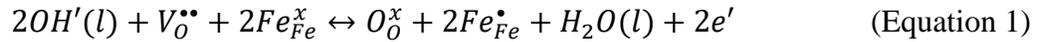

$$2OH'(l) + V_O^{\bullet\bullet} + 2Fe_{Fe}^x \leftrightarrow O_O^x + 2Fe_{Fe}^{\bullet} + H_2O(l) + 2e' \qquad \text{(Equation 1)}$$

According to the Kröger-Vink notation, $OH'$, $Fe_{Fe}^x$, $O_O^x$, and $e'$ represent the OH⁻ ions in the liquid electrolyte, Fe³⁺ and O²⁻ ions in the lattice of the perovskite structure and the electrons generated as a result of oxygen intercalation, respectively. Nevertheless, an alternative pathway based on the incorporation of protons ($OH_O^{\bullet}$) may also explain the changes of Fe⁴⁺, as:[15,31–33]

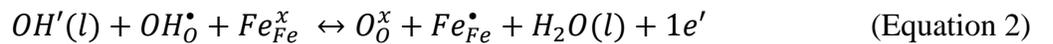

$$OH'(l) + OH_O^{\bullet} + Fe_{Fe}^x \leftrightarrow O_O^x + Fe_{Fe}^{\bullet} + H_2O(l) + 1e' \qquad \text{(Equation 2)}$$

In this case, protons are incorporated in the material through the dissociation of water, reducing the oxidation state of iron as compensation mechanism (please note that protons in the perovskite structure are associated with oxygen atoms with a formal positive charge). This reaction was alternatively proposed for explaining ion intercalation in perovskite oxides.[29] The two mechanisms are schematically illustrated in **Figure 1**.

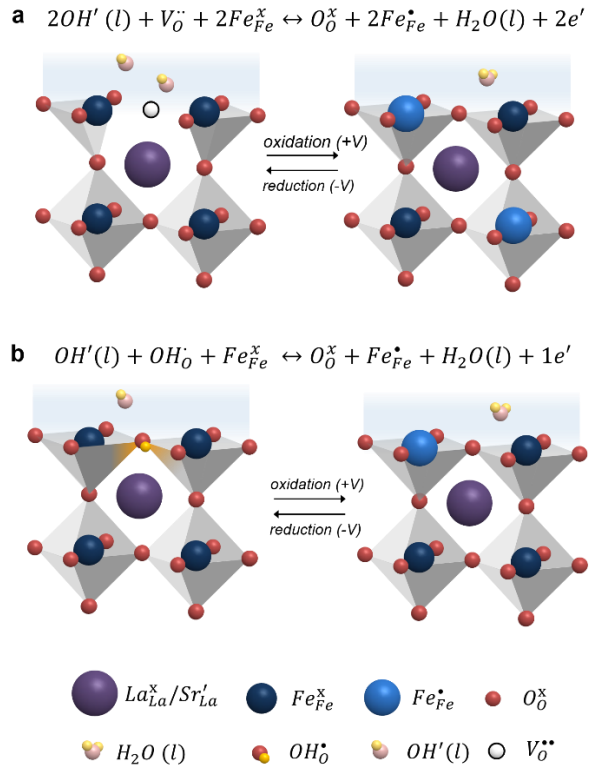

**Figure 1.** Schematic representation of the oxygen (**a**) and protonic (**b**) intercalation mechanisms in alkaline electrolytes.

The present study investigates the mechanism of ion intercalation in LSF50 within liquid alkaline media at room temperature. LSF thin films are employed to ensure a well-defined geometry and an intimate contact with the substrate, which represents an advantage for understanding intrinsic materials properties[34]. However, the use of thin films prevents from employing conventional techniques such as coulometric titration for the determination of the concentration of relevant species (due to a reduced amount of material available). In this work, the concentration of electronic holes during the ion intercalation process was successfully measured by in-situ ellipsometry on thin films. This in-situ ellipsometry methodology has been recently proposed by the authors for successfully tracking defect chemistry and ion insertion in thin films under real electrochemical operation conditions.[35,36] Ellipsometry is a widely employed non-destructive spectroscopic technique, mainly used for the estimation of optical constants and thicknesses, which is based on the change of the polarization state of the light beam reflected on a thin film sample.[35] In combination with ellipsometry, the mechanism of ion intercalation in LSF was investigated using Time-of-Flight Secondary Ion Mass Spectrometry (ToF-SIMS), Positron Annihilation Lifetime Spectroscopy (PALS) and ex-situ ellipsometry. These complementary analyses unambiguously determined that both oxygen and protons are intercalating in LSF in alkaline electrolyte. In addition, a dilute defect chemistry model that reproduces the electronic holes concentration during ion intercalation is proposed.

EXPERIMENTAL SECTION

**Thin film deposition:** $La_{0.8}Sr_{0.2}MnO_{3-\delta}$ (LSM) and LSF50 thin films were sequentially deposited by Pulsed Laser Deposition (PLD) on a 10 mm × 10 mm (100)-oriented Nb doped STO (Nb:STO) substrate (Crystec GmbH) to build up LSF50/LSM/Nb:STO samples. Herein, the LSM interlayer was used for enhancing the electrical conductivity of the sample and creating a polarization in the LSF film.[22,37] LSF layers were deposited using a microfabricated Si mask that allowed the growth of a 5 mm × 5 mm layer in the central area on the top of the LSM. Both layers were deposited with an energy fluency of 0.8 J cm$^{-2}$ at a frequency of 10 Hz using a large-area system from PVD products (PLD-5000) equipped with a KrF-248 nm excimer laser from Lambda Physik (COMPex PRO 205).  The target-to-substrate distance was set to 90 mm and the substrate temperature was kept at 700 ºC. The layers were grown in an oxidizing atmosphere with an oxygen partial pressure of 0.0067 mbar.

**Thin film characterization:** Structural analysis of the thin films was based on high-resolution X-Rays Diffraction (XRD) measurements carried out in θ-2θ Bragg-Brentano geometry using a diffractometer PANAlytical X'Pert-PRO MRD with a Cu $K_\alpha$ radiation source and a combination of a parabolic mirror and a Ge monochromator with two reflections. Scanning Electron Microscopy coupled to Energy-Dispersive X-Ray (SEM-EDX) was carried out in a Zeiss Auriga microscope to characterize the morphology and the elemental composition of the thin films. In this way, the Sr/(La+Sr) ratio for the LSF50 thin film was obtained. The ionic species in the reduced and oxidized samples were investigated by ToF-SIMS depth profiling, using an IONTOF TOF-SIMS 5 instrument. The depth profiling analysis was obtained by alternating a 25 keV Bi+ primary ion and a 2keV Cs+ secondary ion beam. Negative secondary ions were detected in the burst alignment mode. The raster areas were 50 μm x 50 μm and 150 μm x 150 μm for the primary and secondary beams respectively.  The open-volume defects in the post-annealed samples were studied by Variable Energy Positron Annihilation Lifetime Spectroscopy (VEPALS). VEPALS measurements were carried out at the Mono-energetic Positron Source (MePS) beamline at Helmholtz-Zentrum Dresden-Rossendorf (HZDR) in Germany.[38] Digital lifetime CrBr_3 scintillator detector coupled to a Hamamatsu R13089-100 PMT  was utilized for acquisition of positron annihilation signals, which were evaluated using SPDevices ADQ14DC-2X digitizer with 14-bit vertical resolution and 2 GS/s horizontal resolution and with a time resolution function down to about 0.230 ns.[39]

**Electrochemical characterization:** For ex-situ and in-situ experiments in liquid electrolyte, LSF50 and LSM thin films were attached to a copper wire using graphite paste. The frame of the sample and the graphite paste were covered by a robust epoxy resin glue from UHU to avoid undesirable reactions. The final sample presents an active area of 11.7 mm$^2$. The electrochemical measurements were carried out on a three-electrode system in which: i) the encapsulated LSF50 thin film sample was the working electrode (WE); ii) an aqueous 0.1 M KOH solution was employed as an electrolyte; iii) activated carbon and silver wires were introduced into the electrolyte acting as a counter electrode (CE) and reference electrode (RE), respectively. Cyclic Voltammetry (CV) experiments were performed at room temperature in the electrolyte of 0.1 M KOH solution at a scan rate of 0.5 mV s$^{-1}$ using a SP-150 Biologic potentiostat.

**Spectroscopic ellipsometry measurements:** The optical properties of LSF50 thin films were measured with a multi-wavelength spectroscopic ellipsometer (UVISEL, Horiba scientific). The ex-situ ellipsometry experiments were performed in a range of photon energy from 0.6 eV to 5 eV with a step of 0.05 eV whereas the in-situ ellipsometry spectra were recorded in a limited range of photon energy from 1.5 eV to 4.85 eV because of the optical absorption of the liquid electrolyte. The angle of the incident light was set at 70º. The obtained ellipsometry data were processed using the DeltaPsi 2 software from Horiba Scientific. A thickness of 164±2 nm and 78±3 nm was obtained for the LSF50 and LSM thin films, respectively.

For ex-situ measurements, four LSF50/LSM/Nb:STO samples were prepared. On the one hand, two samples were annealed in dry pure oxygen and nitrogen (Nitrogen 5.0) atmospheres at 575 ºC for 5 hours using a Linkam thermal stage (THMS600). These samples were labelled as "O$_2$-annealed" and "N$_2$-annealed", respectively. After cooling down to room temperature, the optical properties of the annealed samples were characterized by spectroscopic ellipsometry in air. On the other hand, two different samples were oxidized and reduced in 0.1 M aqueous KOH electrolyte by applying DC potentials of 0.5 V and -0.4 V in a three-electrode configuration. These samples were labelled as "KOH-oxidized" and "KOH-reduced", respectively.

In case of the in-situ ellipsometry, the measurements were carried out in a homemade 3D printed chamber fabricated in Acrylonitrile Butadiene Styrene (ABS). The chamber was equipped with two transparent optical windows perpendicularly tilted to the incident light beam of the ellipsometer. CE and RE were introduced into the chamber, which was filled with an aqueous electrolyte solution. Finally, LSF50 films under study were polarized by applying DC voltage bias ranging from 0.5 V to -0.4 V in three-electrode configuration. Electrochemical measurements were carried out using a potentiostat (Biologic SP-150). Individual spectroscopic ellipsometry spectra were recorded after each DC voltage step. The end of the intercalation process was ensured by monitoring the electrical current at each DC voltage bias.

RESULTS AND DISCUSSION

**Structural and optical characterization of LSF50:** Fully dense and homogeneous $La_{0.5}Sr_{0.5}FeO_{3-\delta}$ epitaxial thin films of 180 nm in thickness were successfully grown by PLD on $La_{0.8}Sr_{0.2}MnO_{3-\delta}$/Nb:STO(100) substrates (see Supplementary Information **Figure S1**). Interlayers of LSM were included to improve the electrical conductivity of the sample.[22,37] LSF50 films were electrochemically cycled at room temperature in 0.1 M KOH aqueous solution from 0.5 to -0.4 V at a scan rate of 0.5 mV/s (see Experimental Section for more details). **Figure 2a** shows typical Cyclic Voltammetry (CV) curves for LSF50 thin films. In this figure, a pseudocapacitive behavior is clearly observed with redox reaction peaks centered at $E_{1/2}$ = -0.15 V, which is consistent with the results reported in the literature for the similar compound in powder form.[10] To discard contributions from the substrate, similar experiments were carried out for plain LSM/Nb:STO(100) samples and LSF50 layers on top of FTO substrates (see **Figure S2** and **Figure S3** in Supplementary Information, respectively), confirming that the observed redox pair is fully ascribable to the LSF50 layer. The asymmetric peaks observed in the CV scans are likely originated from kinetic limitations, different in cathodic and anodic directions and linked to the sluggish ionic diffusivity in the LSF50 bulk film at room temperature.

In order to explore the origin of the redox reaction peaks, oxidized and reduced LSF50 samples were analyzed by ellipsometry and XRD. These samples were obtained after stabilization in KOH when applying a DC potential of 0.5 V ("KOH-oxidized") and -0.4 V ("KOH-reduced"). **Figure 2b** shows the optical conductivity of both films measured by spectroscopy ellipsometry (details on the data analysis can be found in **Section S4** in Supplementary Information). KOH reduction clearly induces an intensity decrease of low-energy optical transitions (for *A*- and *B*-transitions with photon energies around 1 eV and 3 eV, respectively) while amplification of the one in the high-energy range (*C*-transition with photon energy around 4.2 eV). This behavior has been previously assigned to a strong modification of the LSF band structure, associated with the change in the concentration of $Fe^{4+}$ holes, which induces a redistribution of the optical weight and the creation of new intra-gap states upon hole doping.[35,40,41] To independently confirm this hypothesis, the optical conductivity spectra of LSF50 thin films fully oxidized and reduced in a conventional way (employing oxygen and nitrogen atmospheres at 575 ºC)[35,42,43] were acquired and plotted in **Figure 2b** (named as "$N_2$-annealed" and "$O_2$-annealed", respectively). The similarity of the optical conductivity spectra observed for the LSF layers reduced and oxidized in the different environment confirms that the application of moderate electrochemical potentials in liquid KOH is able to completely oxidize and reduce the LSF.

High-resolution XRD was employed to retrieve the film structural parameters, and to characterize the chemical stability and volume expansion of the $O_2$-annealed, $N_2$-annealed, KOH-reduced and KOH-oxidized samples. The resulting XRD patterns are shown in **Figure 2c**. The whole set of films present a single (00l)-oriented perovskite structure without secondary phase formation and an excellent epitaxial nature. It is important to remark that any of the reduced films show transition to the brownmillerite phase, which is commonly observed for large Sr concentration.[20,22,44,45] The stability of the Nb:STO/LSM substrate is also remarkable as shown in **Figure 2d,** where the diffraction peaks associated to this substrate are always present. A displacement of the diffraction peak position of the $N_2$-annealed LSF50 film from $2\theta$ of 46.75 º to 46.16 º (with respect to the $O_2$-annealed LSF50 film) implies a slight chemical expansion in the out-of-plane lattice parameter of the LSF50 film from 3.883 Å to 3.928 Å ($\approx$2%), which is consistent with the increment in the LSF50 film thickness from 178 nm to 180 nm given by ellipsometry. This expansion is due to the reduction of $Fe^{4+}$ ions to $Fe^{3+}$ ions which have a larger ionic radius as a result of oxygen uptake from the perovskite structure.[40,43,46,47] Comparable peak shift is also observed for KOH-reduced films, which suggests an analogous expansion behavior. Despite this similarity between conventionally- and KOH-reduced samples, a careful analysis of the Kiessig fringes[44,48], which appear as a shoulder on the left side of the (200) diffraction peak of post-annealed LSF50 films while disappear for the KOH-cycled samples, indicates a certain loss of the film-substrate coherence during the ionic intercalation in liquid electrolytes. This is likely due to the foreseen creation of structural defects, as previously referred in the literature.[49,50] Regardless of this decrease of crystalline quality observed by XRD, the microstructural stability of the LSF50 films in 0.1 M KOH solution is proved by SEM analysis of the morphology of the layers after cycling (see **Figure S5** in Supplementary Information), which remains crack-free and fully dense.

Overall, this section showed that the electrochemical cycling in KOH at room temperature leads to a reversible ion intercalation, resulting in a $Fe^{4+}/Fe^{3+}$ transition as a charge compensation mechanism.

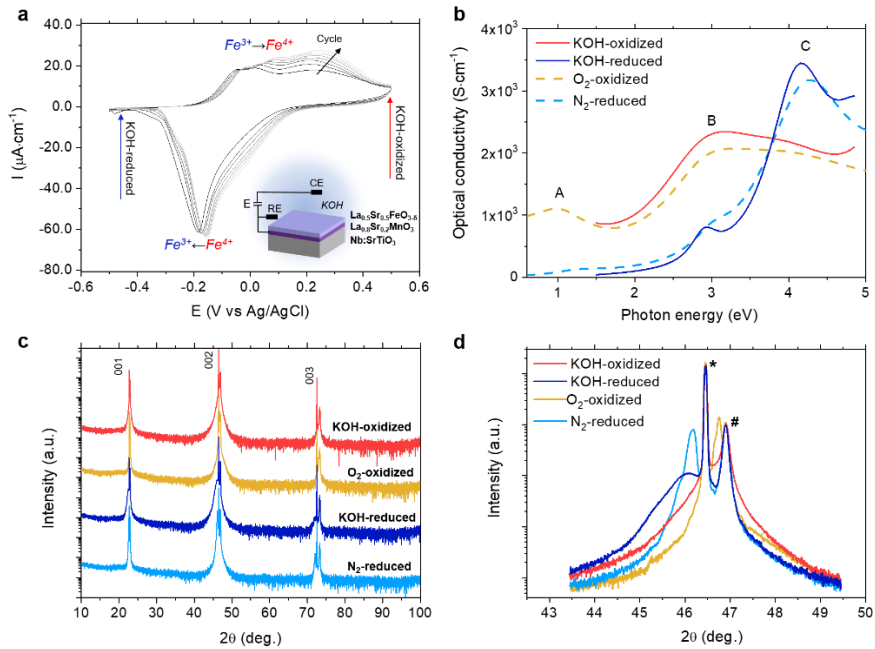

**Figure 2. a.** CV curve of the LSF50 thin film measured at room temperature using a scan rate of 0.5 mV/s. **b.** Optical conductivity spectra of the post-annealed LSF50 films (dash lines), KOH-reduced and KOH-oxidized LSF50 films (solid lines) obtained by ex-situ ellipsometry. **c.** XRD diagrams of the $O_2$-annealed, $N_2$-annealed, KOH-reduced and KOH-oxidized LSF50 samples. **d.** Magnification of the (002) diffraction peak. The diffraction peaks of Nb:STO and LSM are marked with star (*) and hash (#) symbols, respectively. The observed shoulder on annealed samples correspond to Kiesseg fringes characteristic of coherent and homogeneous epitaxial layers.

**In-situ ellipsometry of LSF50 redox cycling in alkaline electrolytes:** A high sensitivity of the LSF optical properties to the concentration of electronic holes was reported in the previous section. Here, in-situ ellipsometry acquired during the electrochemical reduction is employed for getting more insights into the mechanism of ion intercalation. The measurements were carried out in a three-electrode sample using a homemade chamber (see Experimental Section). In order to measure the real equilibrium conditions, we focused on the reduction reaction at room temperature applying DC voltage to the sample in stepwise from 0.5 V to -0.4 V, see **Figure 3a**. The equilibrium was considered achieved when the electrochemical current, measured after each polarization step, reached a value close to zero. The extracted optical conductivity spectra of the LSF50 thin film measured at various intercalation potentials are shown in **Figure 3b** (see **Figure S6** in Supplementary Information for the measured ellipsometry raw data and **Section S4** in Supplementary Information for more details about the fitting model). The data shows a monotonic variation in the intensity of the optical transitions "A", "B" and "C", consisting in a weakening of low energy transition "A" and "B" together with a strengthening of UV transition "C" with cathodic potential (*i.e.* with the reduction reaction). This trend implies a decrease of the hole concentration in the LSF layer.[40,41,51,52]

In our previous work, the concentration of electronic holes in LSF has been directly related to the intensity of the low-energy transition "A".[35] However, the high optical absorption of the liquid electrolyte in the infrared region hinders the application of the same procedure in this study. In order to overcome such a limitation, we focus here on the variation of the LSF films' optical conductivity at the photon energy of 2 eV. In supplementary information **Section S7,** we demonstrate that a linear relation with the $Fe^{4+}$ holes concentration is present. As a result, the $Fe^{4+}$ holes concentration at different intercalation potentials can be readily quantified, see **Figure 3c**.

The curve suggests a single electrochemical exchange process (proton intercalation or oxygen extraction) during the reduction, as shown by the monotonic decrease of the iron valence. A pseudo-plateau for the Fe valence is observed around -0.05 V. The maximum capacity that can be extracted from the ionic intercalation in the LSF layers corresponds to the full reduction of the Fe valence from 4+ to 3+ and is determined by the total concentration of $Sr'_{La}$ dopant (around 55 mAh/g in the LSF50 film under study).

It is also interesting to compare the charge transferred after each voltage step (*i.e.* incremental capacity[53]) calculated by ellipsometry and by electrochemical method. Since the $Fe^{4+}/Fe^{3+}$ transition is accompanied by charge transfer as:

$$Fe^{\bullet}_{Fe} + 1e' \rightarrow Fe^{x}_{Fe} \qquad \text{(Equation 3)}$$

The number of charge transferred in the reduction reaction can be calculated via the ellipsometry approach ($Q_{elliposmetry}$) as:

$$Q_{elliposmetry} = n \cdot [Fe^{\bullet}_{Fe}] \cdot F \qquad \text{(Equation 4)}$$

Where *n* is the number of LSF50 cell (in mol) in the thin film and *F* is the Faraday constant. On the other hand, the charge transferred at each stepwise voltage can be obtained calculating the area under the I-t curves of **Figure 3a**, as:[54]

$$Q_{electrochemical} = \int_{t1}^{t2} I(t)dt \qquad \text{(Equation 5)}$$

**Figure 3d** shows the incremental capacity obtained from the electrochemical and ellipsometry measurements employing **Equations 4** and **5**, respectively. A good agreement in the tendency of the incremental capacity obtained by the two approaches is observed, confirming the capability of ellipsometry to track the concentration of $Fe^{4+}$ in the layer. Nevertheless, the electrochemical method shows a systematic larger value of charge transferred at each voltage step. This behavior can be ascribed to the electrochemical reactions (charge transfer) on the film surface caused, for instance, by oxygen reduction reaction (ORR).[25,55] Since in-situ ellipsometry is bulk-sensitive, it is able to differentiate the battery-like behavior linked to the ionic insertion from the electrochemical reactions taking place on the sample surface.[36] It is important to note that different fitting models were considered to fit the ellipsometry spectra and to derive the hole concentration at each potentials (see **Section S8** in Supplementary Information). Although a certain discrepancy in the thickness of the LSF layer was observed across the models, all of them led to the same concentration of $Fe^{4+}$, validating the presented in-situ ellipsometry approach.

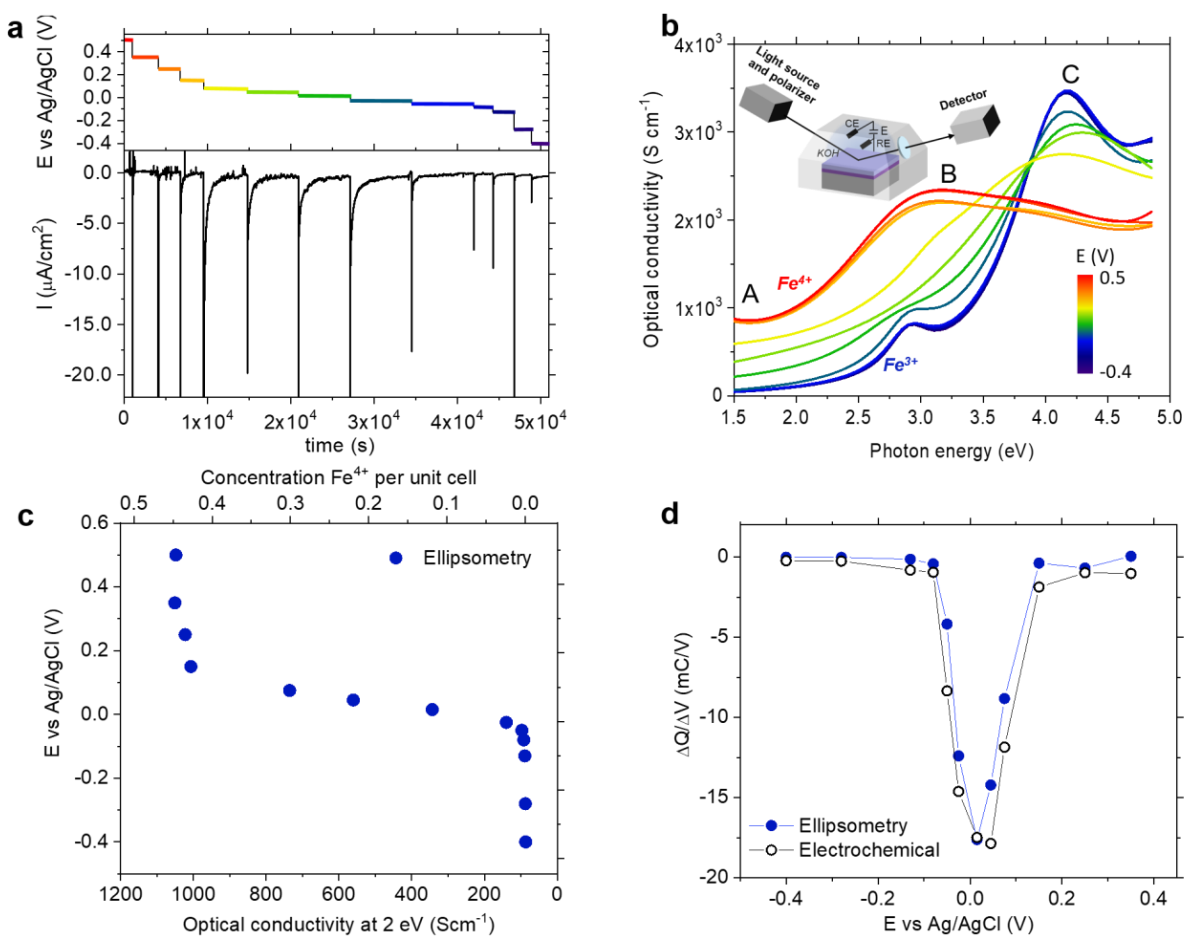

**Figure 3. a.** Evolution of the applied electrochemical potentials and of the current measured over time in the in-situ experiment. **b.** Optical conductivity spectra of the LSF50 thin film were recorded at various intercalation potentials (applied voltages). The inset shows a schematic representation

of the chamber used for the measurements. **c**. Optical conductivity measured at 2 eV and corresponding $Fe^{4+}$ concentration measured at each intercalation potential. **d**. Comparison of the incremental capacity of the LSF50 thin film obtained from the electrochemical and ellipsometry results.

**Ion intercalation mechanism in LSF50 cycled in an alkaline electrolyte:** As mentioned in the introductory section (and sketched in **Figure 1**), proton and oxide-ion incorporation pathways may be active separate or together. To unambiguously determine the active incorporation pathway, the redox process of LSF in liquid electrolyte was investigated by ToF-SIMS, PALS, and ex-situ ellipsometry measurements.

The possible proton incorporation was studied by ToF-SIMS in the KOH reduced sample, where, according to the protonic pathway shown in **Equation 2**, protons should compensate the reduction of oxidation state of iron atoms. **Figure 4a** shows the depth profiles of the $H^-$ and $OH^-$ intensities measured for the KOH-reduced sample. The thickness of the LSF, LSM and STO layers was obtained by monitoring the $FeO^-$, $MnO^-$ and $TiO^-$ signals, respectively (**Figure S9** in Supplementary Information). An increased signal of the $H^-$ and $OH^-$ is visible in the LSF50 layer with respect to the background signal (STO layer). This experiment undoubtedly shows that an enhanced proton concentration is present is LSF as a consequence of incorporation during the electrochemical reduction in KOH electrolyte, proving that the protonic pathway is active. The sharp decrease of hydrogen species at the LSF/LSM interface suggests that, within the electrochemical window studied, protons are not intercalated in the LSM layer.

The oxygen intercalation mechanism was studied by ex-situ ellipsometry. Since oxygen vacancies cannot be straightforwardly measured in thin film oxides, we performed an indirect measurement to verify the presence of oxygen intercalation in the LSF50 thin film. First, a LSF50 sample was annealed in dry $N_2$ at T=575 °C. As shown in the optical conductivity of **Figure 4b**, the $N_2$-annealed LSF50 film was fully reduced. Since a dry atmosphere was used, no protons are expected in this state and the dopant is entirely compensated by oxygen vacancies ($[Fe^{\cdot}_{Fe}]\sim0$, $2[V^{\cdot\cdot}_O]\sim[Sr'_{La}]$). The sample was then oxidized in 0.1 M KOH electrolyte by applying potentials of 0.5 V. After this treatment, the optical measurement shows that an almost complete oxidation is achieved in the layer, highlighted by the large increase of optical conductivity in the low energy region, see **Figure 4b**. According to the protonic pathway, positive bias should decrease the concentration of protons in the layers. Since no protons were originally present in the sample, the only possible explanation for the oxidation of the layer is that the oxygen pathway is active and oxygen can enter in the reduced sample filling the oxygen vacancies and increasing the oxidation state of Fe (see **Equation 1**).

An analogous experiment was also performed with a sample completely oxidized in dry $O_2$ atmosphere (**Figure S10** in Supplementary Information). Also in this case, complete oxidation and reduction was observed in the sample after electrochemical bias in KOH. It is also interesting to notice that the CV of the two differently annealed samples (**Figure S10** in Supplementary Information) are comparable and present the redox peak at the same voltage, despite starting from a very different oxygen vacancy (and electronic holes) concentration. This also suggests that a new equilibrium is achieved in liquid electrolytes, in which both oxygen and protonic species are intercalating in the layer.

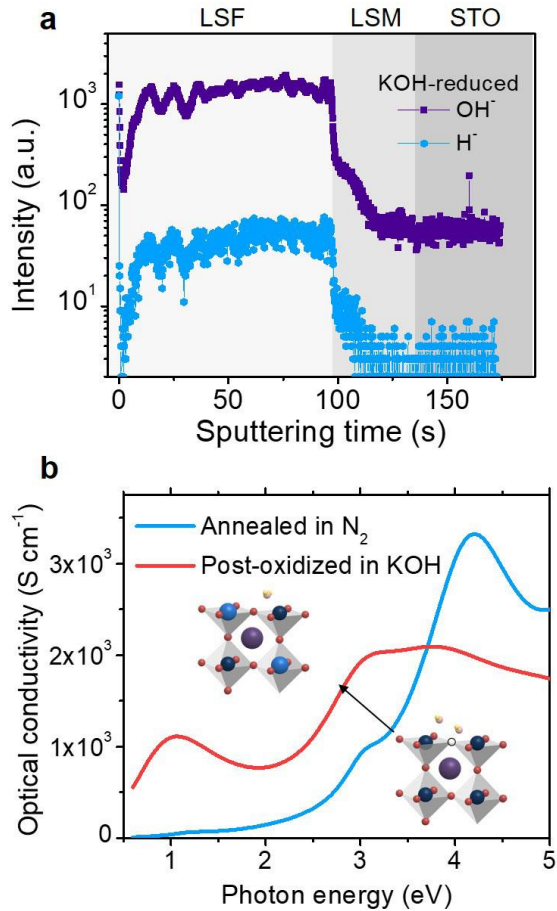

**Figure 4. a**. ToF-SIMS depth profiles of H⁻ and OH⁻ ions for the KOH-reduced sample. A clear increase of the protonic species intensity is observed in the LSF50 top-layer. **b**. Optical conductivity of a LSF50 layer reduced in dry $N_2$ followed by an oxidation in KOH. The increase of optical conductivity shows that the oxygen mechanism is also active in alkaline electrolytes.

The LSF50 thin films subjected to the different redox treatments were also studied by Variable Energy Positron Annihilation Lifetime Spectroscopy (VEPALS) for direct analysis of defect type and concentration. PALS is particularly sensitive to localized vacancy-like defects (or close porosities) of a neutral or negative charge, which acts as annihilation sites for positrons. The positron lifetime is proportional to the size of the annihilating defect. By varying the positron implantation energy ($E_p$), which is related to the mean positron implantation depth in the sample, one can obtain $z$-resolved information on defect type (from the positron lifetime $\tau_i$) and relative concentration (from the relative intensity $I_i$) with nominal nm-resolution (notwithstanding a certain signal broadening for increasing $z$). The PALS analysis shown below is based on dedicated studies on a related perovskite, $SrTiO_3$ (STO), for the identification of the defect type based on the positron lifetime.[56–58] Although PALS cannot directly measure positive defects such as oxygen vacancies or protons, the technique is sensitive to their association with negative defects (e.g. cationic vacancies) resulting in the formation of an associate. A higher oxygen vacancy content is expected to lead to an increase of defect lifetime owing to the formation of larger open clusters.[59,60] Conversely, the effect of proton uptake depends on the defect site: the association of protons with a negatively charged vacancy leads to a contraction of the open volume and to lifetime decrease for proton incorporation as interstitials,[61] whereas protons bonded to oxygen may give rise to a displacement of oxygen from the equilibrium sites, resulting in a distorted structure and larger defect open volumes.[17,49] . **Figure 5a** shows the average defects size positron lifetime ($\tau_{av}$) obtained as a function of $E_p$ for the KOH-oxidized and reduced samples as well as the $O_2$-oxidized and the $N_2$-reduced LSF50 thin films. Here, $\tau_{av} = \Sigma_i \tau_i \cdot I_i$, as obtained from deconvolution of the measured lifetime in two-lifetime components (cf. also experimental). The average defect size ($\approx$200 ps) obtained in the LSF50 film falls into the range of B-site vacancy.[58] The $N_2$-annealed and KOH-reduced samples are characterized by a larger average positron lifetime $\tau_{av}$ with respect to the oxidized counterpart ($O_2$-annealed and KOH-oxidized, respectively), i.e., the reduction process induces the formation of larger defect clusters – arguably by increasing the concentration of oxygen vacancies. Interestingly, for the KOH-oxidized and reduced layers, $\tau_{av}$ is larger than for the $O_2$-annealed and $N_2$-annealed samples, respectively, suggesting that a more defective structure is induced upon treatment in liquid. All layers present similar average lifetimes for the LSM interlayer, confirming that the different treatments do not affect this material. The subsurface layer ($z$ <20 nm) is characterized by remarkably higher $\tau_{av}$, indicating a large increase in defect size and concentration, which is related to the surface annihilation sites.

An analysis of the single lifetime components ($\tau_{d1}$ and $\tau_{d2}$) and the relative intensities allow obtaining further insights, see **Figure 5b**. Comparing with literature values of STO,[56] the dominant defect component $\tau_{d1}$ in the LSF50 film ($I_{d1}$ >60 %, $I_{d1}$+ $I_{d2}$= 100%) is found to be in the range of B-site vacancy ($V_{Ti}''''$) and a complex between Ti and oxygen vacancies (consistently with the average lifetime analysis, cf. panel a). Reducing the sample (both in KOH and in $N_2$) is observed to increase the major defect lifetime $\tau_{d1}$, indicating the formation of larger $V_{Ti}''''$ - $V_O^{\cdot\cdot}$ associates. The minority lifetime component $\tau_{d2}$ is instead associated to defect clusters containing Sr-vacancies ($V_{Sr}''$), such as $V_{Sr}''+V_O^{\cdot\cdot}$ or full Schottky defects ($V_{Sr}'' + V_{Ti}'''' + 3V_O^{\cdot\cdot}$). $I_{d2}$ shows a considerable increment for the KOH-reduced sample (note that $I_{d1}$ decreases accordingly). Here, an additional effect of proton uptake on lattice distortion and consequent increase of defect volume may be present. This hypothesis appears to be supported by the decrease of crystalline quality observed in the XRD experiments (**Figure 2d**), possibly related to hydration-induced lattice distortions and LSF50 structure disorder.

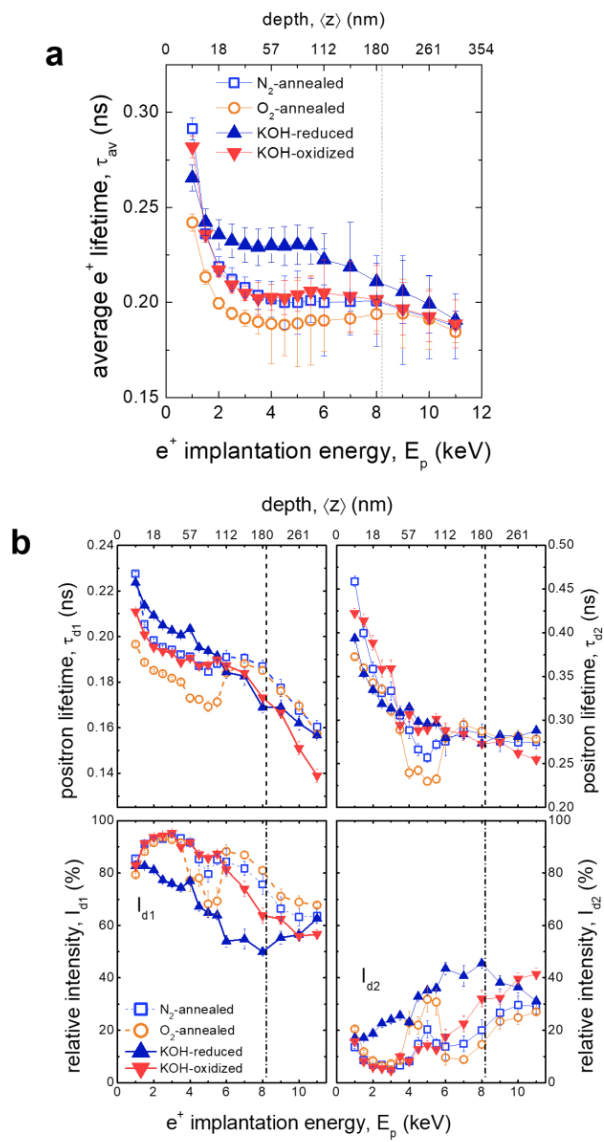

**Figure 5.** PALS parameters analysis for the $O_2$-annealed, $N_2$-annealed, KOH-reduced and KOH-oxidized samples. **a.** Average positron lifetime $\tau_{av}$. **b.** The first $\tau_{d1}$ (left) and the second $\tau_{d2}$ (right) components of the positron lifetime and their relative intensities $I_{d1}$ and $I_{d2}$. The vertical line at $E_p=8$ keV shows the position of the interface between LSF and LSM films.

In summary, according to the ex-situ measurements with different techniques, both proton and oxygen ions are intercalated into the LSF50 layers during the electrochemical experiments in alkaline electrolyte, giving rise to a variation of $Fe^{4+}$ holes for electro-neutrality. At cathodic (negative) intercalation potentials, the electrical field causes oxygen anions to be pumped out from the LSF50 structure and protons from the liquid electrolyte to be incorporated into the LSF50 structure, while the opposite reactions take place at anodic potentials.

**Defect chemistry modelling:** The results obtained in the previous sections suggest that: (i) Protons enter the LSF50 thin films upon electrochemical reduction. (ii) Oxygen is also exchanged during ionic intercalation with the electrolyte and the same equilibrium can be reached regardless of the initial defect concentration in the layer. (iii) A single electrochemical process is observed upon reduction, i.e. protons and oxygen are exchanged together. Therefore, ion intercalation can be described as a combination of oxygenation and hydrogenation reactions involving oxygen anions and protons. The global reaction can be written as a sum of the oxygenation reaction (**Equation 1**) and the hydration reaction (**Equation 2**) as:

$$3OH_l' + V_O^{\bullet\bullet} + OH_O^{\bullet} + 3Fe_{Fe}^x \leftrightarrow 2O_O^x + 3Fe_{Fe}^{\bullet} + 2H_2O(l) + 3e_l' \qquad \text{(Equation 6)}$$

Applying the Nernst equation, the potential for the ion intercalation for dilute non-interacting defects can be expressed as:

$$E = E_0 + \frac{RT}{3F}\ln\left(\frac{[O_O^x]^2[Fe_{Fe}^{\bullet}]^3}{[V_O^{\bullet\bullet}][OH_O^{\bullet}][Fe_{Fe}^x]^3}\right) + \frac{RT}{3F}\ln\left(\frac{[H_2O]^2}{[OH']^3}\right) \qquad \text{(Equation 7)}$$

Where $E_0$ is the standard potential of the ionic intercalation, $R$ is the gas constant (8.314 Jmol$^{-1}$K$^{-1}$). The first logarithmic component on the right-hand side of the equation is linked to the concentration of the point defects in the LSF50 thin film, and the second logarithmic term is associated with the concentration of the ions in the liquid electrolyte with the activity coefficient assumed to be 1.[62,63] Thus, **Equation 7** can be rewritten as:

$$E = E_0 + \frac{RT}{3F} \ln\left(\frac{[O_O^x]^2 [Fe_{Fe}^\bullet]^3}{[V_O^{\bullet\bullet}][OH_O^\bullet][Fe_{Fe}^x]^3}\right) + 0.16\, V \qquad \text{(Equation 8)}$$

Because the electronic transition $Fe^{3+}/Fe^{2+}$ was not detected in the studied potential window,[10] the $Fe^{2+}$ electrons in the LSF50 are negligible. Thus, the charge equilibrium in the LSF50 film is:[43,64,65]

$$2[V_O^{\bullet\bullet}] + [OH_O^\bullet] + [Fe_{Fe}^\bullet] = [Sr_{La}'] \qquad \text{(Equation 9)}$$

$$[Fe_{Fe}^\bullet] + [Fe_{Fe}^x] = 1 \qquad \text{(Equation 10)}$$

The ratio between the concentration of oxygen vacancies $V_O^{\bullet\bullet}$ and $OH^\bullet$ ions can be determined by the hydration reaction:[63,66,67]

$$H_2O(l) + V_O^{\bullet\bullet} + O_O^x \leftrightarrow 2OH_O^\bullet \qquad \text{(Equation 11)}$$

With the equilibrium constant $K_w$:

$$K_w = \frac{[OH_O^\bullet]^2}{[H_2O_l][V_O^{\bullet\bullet}][O_O^x]} \qquad \text{(Equation 12)}$$

The hydration equilibrium constant depends on the material´s properties and defines the total amount of protons that can be incorporated into the layer. The protonation mechanism of ferrites perovskites in high-temperature humid environment was studied in detail by Prof. Maier and co-workers.[32,68–71] Interestingly, they found that in the intermediate temperature range 250 °C - 500 °C LSF presents quite a low protonic concentration, especially when compared with Ba-doped ferrites. Nevertheless, the protonation mechanism (**Equation 11**) is regulated by an endothermic reaction, meaning that the concentration of protons will be higher at lower temperatures. Considering the enthalpy and entropy of the hydration reaction measured for LSF,[32,68–71] a value of $K_w$ around 0.1 can be extrapolated at room temperature, which is comparable with the protonation constant measured for a good protonic conductor such as $Ba_{0.5}Sr_{0.5}Fe_{0.8}Zn_{0.2}O_{3-\delta}$ at 400 °C. Because of Mattrix effects, no quantitative information on the concentration of protons in the reduced LSF50 thin films can be deduced from ToF-SIMS measurements. Thus, a value of $K_w$=0.1 has been considered in this study. The set of **Equations 7-12** can be combined for obtaining an expression where the electrochemical potential of intercalation can be described as a function of $Fe^{4+}$ holes concentration, as:

$$E = E_0 + \frac{RT}{3F}\ln\left(\frac{(2.55+[V_O^{\bullet\bullet}]+[Fe_{Fe}^{\bullet}])^2\,[Fe_{Fe}^{\bullet}]^3}{[V_O^{\bullet\bullet}](0.45-2[V_O^{\bullet\bullet}]-[Fe_{Fe}^{\bullet}])(1-[Fe_{Fe}^{\bullet}])^3}\right) + 0.16\,V \qquad \text{(Equation 13)}$$

Where $[V_O^{\bullet\bullet}]$ is the solution of the second-order equation:

$$2.55 + [V_O^{\bullet\bullet}] + [Fe_{Fe}^{\bullet}] = \frac{(0.45-2[V_O^{\bullet\bullet}]-[Fe_{Fe}^{\bullet}])^2}{K_w[V_O^{\bullet\bullet}]} \qquad \text{(Equation 14)}$$

With these equations, it is possible to fit the equilibrium $Fe^{4+}$ concentration obtained at each electrochemical potential considering only $E_0$ as a free parameter. **Figure 6** shows the results of the model obtained for a $E_0 = 0.83$ V vs RHE (or -0.083 V vs Hg/HgO). An excellent fit of the experimental data is obtained, suggesting that the proposed model is able to describe the mechanism of ionic intercalation in LSF50 thin films.

The model shows that the concentration of protons and oxygen vacancies increases while reducing the samples until the majority of holes is consumed. These results demonstrate that the ionic intercalation in LSF is dominated by bulk defect chemistry, which defines the potential and shape of the electrochemical response. In this sense, it is worth noting that the intercalation potential $E_0$ is related to the oxidation constant commonly measured in these oxides at high temperature gaseous environment, which describes the equilibrium concentration of defects as a function of the external oxygen chemical potential (varied either by the external oxygen partial pressure or by applying an electrochemical potential in a solid oxide cell).[35,72] Depending on the reducibility of the materials, the intercalation potential window in both liquid and gaseous environment is strongly modified. Finally, the effect of the hydration reaction on the defect chemistry of perovskite oxides in liquid electrolytes should be discussed. Depending on the capability of a certain material to host protons (*i.e.* on the value of the hydration constant $K_w$), the oxygen or protonic electrochemical exchange mechanism will be dominating, giving rise to significant changes in the $V_O^{\bullet\bullet}$ / $OH^\bullet$ ratio in reducing conditions.

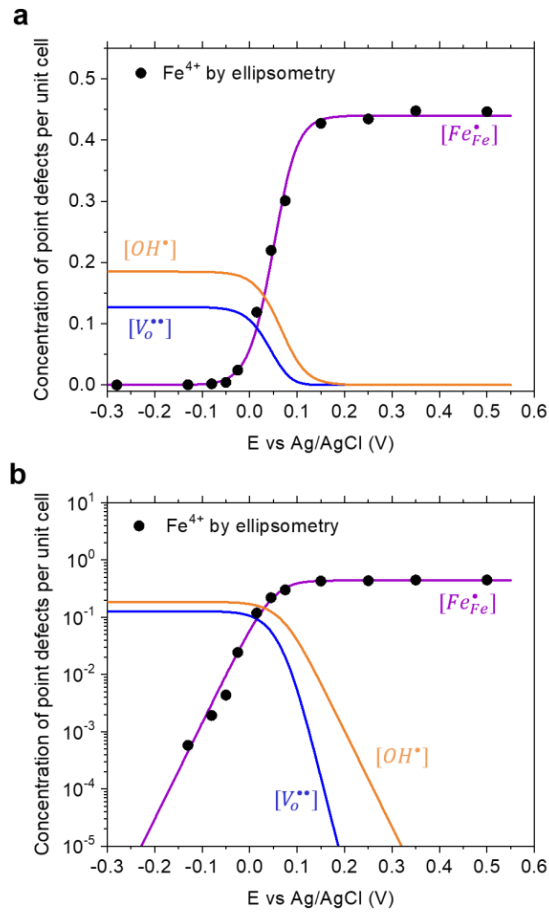

**Figure 6.** Linear (**a**) and logarithmic (**b**) plots of the defects concentration obtained by defect chemistry modelling (lines) and of the $Fe^{4+}$ holes concentration measured by ellipsometry (points) for LSF50 thin films in 0.1 M KOH at room temperature.

CONCLUSIONS

Ion intercalation phenomena for LSF50 thin films in liquid alkaline electrolyte were studied in this paper. One can conclude that, under an electrochemical field (induced by an applied voltage), ions from the liquid electrolyte intercalate into the LSF50 resulting in the $Fe^{4+}/Fe^{3+}$ electronic transition. The intercalation mechanism was found to consist in a mixed oxygenation and hydration reactions (as determined by ex-situ XRD, ex-situ ellipsometry, SIMS and PALS techniques). It was possible to derive the concentration of $Fe^{4+}$ holes in the LSF50 film from the optical conductivity via the linear proportionality of the optical conductivity at the photon energy of 2 eV to the holes concentration. Like this, the evolution of $Fe^{4+}$ concentration with intercalation potential was satisfactorily described by a dilute defect chemistry model, which is highlighted by the importance of the oxygenation and hydration reactions in the intercalation mechanism. Overall, in-situ/operando ellipsometry approach got new insights into the defect chemistry of LSF50 oxide thin films along the ion intercalation in alkaline media, opening a new way to tailor the LSF50 thin film's functional properties for energy and information applications.

ASSOCIATED CONTENT

**Supporting Information**. The following Supporting Information is available free of charge via the Internet at https://pubs.acs.org/.

XRD diagrams and SEM image of the as-deposited LSF50/LSM/Nb:STO sample, CV cycle for the LSF50 thin film grown on FTO substrates, in-situ ellipsometry measurements for LSM thin films, ellipsometry data analysis, SEM images of the morphology of the post-treated LSF50 thin films, ellipsometry raw data as a function of intercalation potential, linear relation between the optical conductivity and concentration of electron holes, models used for modelling the ellipsometry data, ToF-SIMS depth profiles of the ions in the KOH-reduced sample, ex-situ ellipsometry measurements. (PDF)

AUTHOR INFORMATION

**Corresponding Author**

*Francesco Chiabrera  -Department of Advanced Materials for Energy Applications, Catalonia Institute for Energy Research (IREC), Jardins de les Dones de Negre 1, 08930, Sant Adrià del Besòs, Barcelona, Spain.

- Department of Energy Conversion and Storage, Functional Oxides group, Technical University of Denmark, Fysikvej, 310, 233, 2800 Kgs. Lyngby, Denmark.

Email: fmach@dtu.dk;

 *Albert Tarancón -Department of Advanced Materials for Energy Applications, Catalonia Institute for Energy Research (IREC), Jardins de les Dones de Negre 1, 08930, Sant Adrià del Besòs, Barcelona, Spain.

 -ICREA, Passeig Lluís Companys 23, 08010, Barcelona, Spain.

 Email: atarancon@irec.cat

**Author Contributions**

The manuscript was written through contributions of all authors. All authors have given approval to the final version of the manuscript.

ACKNOWLEDGMENT

This research was supported by the funding from the European Research Council (ERC) under the European Union's Horizon 2020 research and innovation programme (ULTRASOFC, Grant Agreement number: 681146) and the funding from the NANOEN project (2017 SGR 1421). This research was also supported by the funding from the European Union's Horizon 2020 research and innovation program under grant agreement No 824072 (HARVESTORE). Parts of this research were carried out at ELBE at the Helmholtz-Zentrum Dresden - Rossendorf e. V., a member of the Helmholtz Association. We would like to thank the facility staff (Eric Hirschmann and Ahmed G. Attallah) for assistance.